\documentclass[prl,twocolumn,aps,floats,floatfix,superscriptaddress]{revtex4-1}

\usepackage{bbold}
\usepackage{bbm}
\usepackage[pdftex]{graphicx}
\usepackage{latexsym,amsmath,verbatim}
\usepackage{color}
\usepackage{rotating}
\usepackage{multirow}
\usepackage[normalem]{ulem}
\usepackage[english]{babel}

\begin{document}
\title{Dynamical leaps due to microscopic changes in multiplex networks}

\author{Marina Diakonova}
\affiliation{School of Mathematical Sciences, Queen Mary University of London, London E1 4NS, United Kingdom}
\affiliation{Instituto de F\'{\i}sica Interdisciplinar y Sistemas Complejos IFISC (CSIC-UIB), Campus UIB, 07122 Palma de Mallorca, Spain}

\author{Jos\'e J. Ramasco}
\author{V\'{\i}ctor M. Egu\'iluz}
\affiliation{Instituto de F\'{\i}sica Interdisciplinar y Sistemas Complejos IFISC (CSIC-UIB), Campus UIB, 07122 Palma de Mallorca, Spain}

\date{\today}

\begin{abstract}
Recent developments of the multiplex paradigm included efforts to understand the role played by the presence of several layers on the dynamics of processes running on these networks. The possible existence of new phenomena associated to the richer topology has been discussed and examples of these differences have been systematically searched. Here, we show that the interconnectivity of the layers may have an important impact on the speed of the dynamics run in the network and that microscopic changes such as the addition of one single inter-layer link can notably affect the arrival at a global stationary state. As a practical verification, these results obtained with spectral techniques are confirmed with a Kuramoto dynamics for which the synchronization consistently delays after the addition of single inter-layer links.   
\end{abstract}


\maketitle

The ubiquity of processes naturally described by dynamics on networks raises important issues on the role played by network structure on the emergence of collective phenomena. It has been shown that spectral analysis of the associated adjacency and Laplacian matrices can offer insights into a variety of fundamental phenomena such as those relying upon spreading or diffusion mechanisms \cite{Boguna2003, Jamakovic2006, Dorogovtsev2008}. In particular, spectral methods are the basis to characterize synchronization and random walk diffusion in networks \cite{Atay2006, Arenas2006, GG2007, Chen2012}. The second smallest eigenvalue $\lambda_2$ is known to be related to the timescale to synchronization \cite{Almendral2007} and consensus \cite{Estrada2015}, and  is often interpreted as the `proper time' of the system to relax \cite{Hernandez2014}. This quantity, which depending on the literature is known as `algebraic connectivity' or `spectral gap', is also indicative of the  time of diffusion \cite{Lovasz1994}. While the role of $\lambda_2$ in ``simple" graphs is well understood, more effort is needed to characterize its role and behavior in more realistic (and therefore complex) contexts.

One such novel framework is that of a multilayer network \cite{Boccaletti14, Kivela14, Goh2015}. Its usefulness extends from finance \cite{Iasio2013, Thurner2015, Burkholz2015} and mobility \cite{Domenico2014,Strano2015}, to epidemics \cite{Granell2013,Vazquez2015} and societal dynamics \cite{Mucha10, Buldyrev2010, Klimek2013, Csermely2013, Wang2013, Diakonova2014,Klimek2015}. The multiplex scenario is ostensibly non-trivial, in the sense that the phenomena observed on this system of interconnected networks cannot be straightforwardly reduced to an aggregate network \cite{Diakonova2015}. The implication is that the multiplex structure plays a fundamental role in diffusive processes, and that
therefore its effect of on $\lambda_2$ is a pertinent and open issue.

Diffusion on a multiplex, that is, a system with layers of $N$ nodes with an association of corresponding nodes in each layer, were considered in \cite{Gomez2013, SoleRibalta2013}. By varying the inter-layer diffusion constant, these works found boundaries of $\lambda_2$ for the multiplex in terms of the values for individual layers (see also \cite{Sahneh2015}). For a general 
weighted two-layer multiplex with a varying inter-layer link weight $p$, these results were rephrased in terms of a structural transition \cite{Radicchi2013}:
it was found that below some critical $p_c$, $\lambda_2(p) \sim 2\, p$, while for larger $p$ the algebraic connectivity approaches an asymptote given by $L_2 = \frac{1}{2}\, \lambda_2 \,\left( \mathcal{L}_1 +  \mathcal{L}_2 \right)$,
where $\mathcal{L}_{1,2}$ are the Laplacian matrices of the two layers. The changing character of the functional growth of $\lambda_2$ suggests existence of two regimes, an `underconnected' one with the two layers functioning autonomously, and a `multiplex' one where the inter-layer connectivity plays the dominant role. It has been noted that the existence of the point of inflection $p_c$ might follow from linear algebra arguments \cite{Garrahan2014}. At this point it is also not certain whether $p_c$ tends to zero in the limit of $N \rightarrow \infty$, or whether it is indeed significant in describing
the observed changes in the dynamic timescales \cite{Bastas2015}.

This work extends the analysis of multiplex timescales to cover the effect of the specific type of inter-layer coupling. Instead of gradually tuning up the intensity of inter-layer links $p$, we now switch on the inter-layer links at unit intensity one by one. Our strategy of considering binary weight on the inter-layer links corresponds to a natural situation where these links are either absent or present, and where one can preferentially control the amount, and the placement, of such links. Spectral properties resulting from such inter-layer link sequences, and their placement, were considered in \cite{Hernandez2014}. The authors studied identical partially-coupled networks, and found that the trends of average algebraic connectivity showed a qualitative change at an (analytically derivable) minimum number of inter-layer links, depending on the placement of their end points. The authors of \cite{Shakeri2015} elucidated the optimal relation between the similarity of layers and the weight of inter-layer link, given some minimization constraint. Here we consider non-identical, general layers (see \cite{Gambuzza2015} for study of a multiplex with widely different inter-layer structure), and instead focus on the effect of the addition of inter-layer links to individual systems. We find that $\lambda_2(q)$ of the fraction $q$ of such an inter-layer connection sequence is characterized by leaps, implying sudden decreases in the timescales of the resulting multiplex. We characterize the statistics of these jumps and elucidate the way in which layer degree correlation \cite{Radicchi2014,Enzo2015} affect them. Our results show that not only are highly correlated multiplexes connected with high intensities on average slower than anti-correlated multiplexes, but that the statistics of jumps are qualitatively different for distinct types of ordering. Finally, we validate our findings by running Kuramoto dynamics on the multiplex, and verifying that $\lambda_2$ does indeed inform on the scaling of the approach to the fully-synchronized state.

\begin{figure}
\centering
\includegraphics[width = 8.6cm]{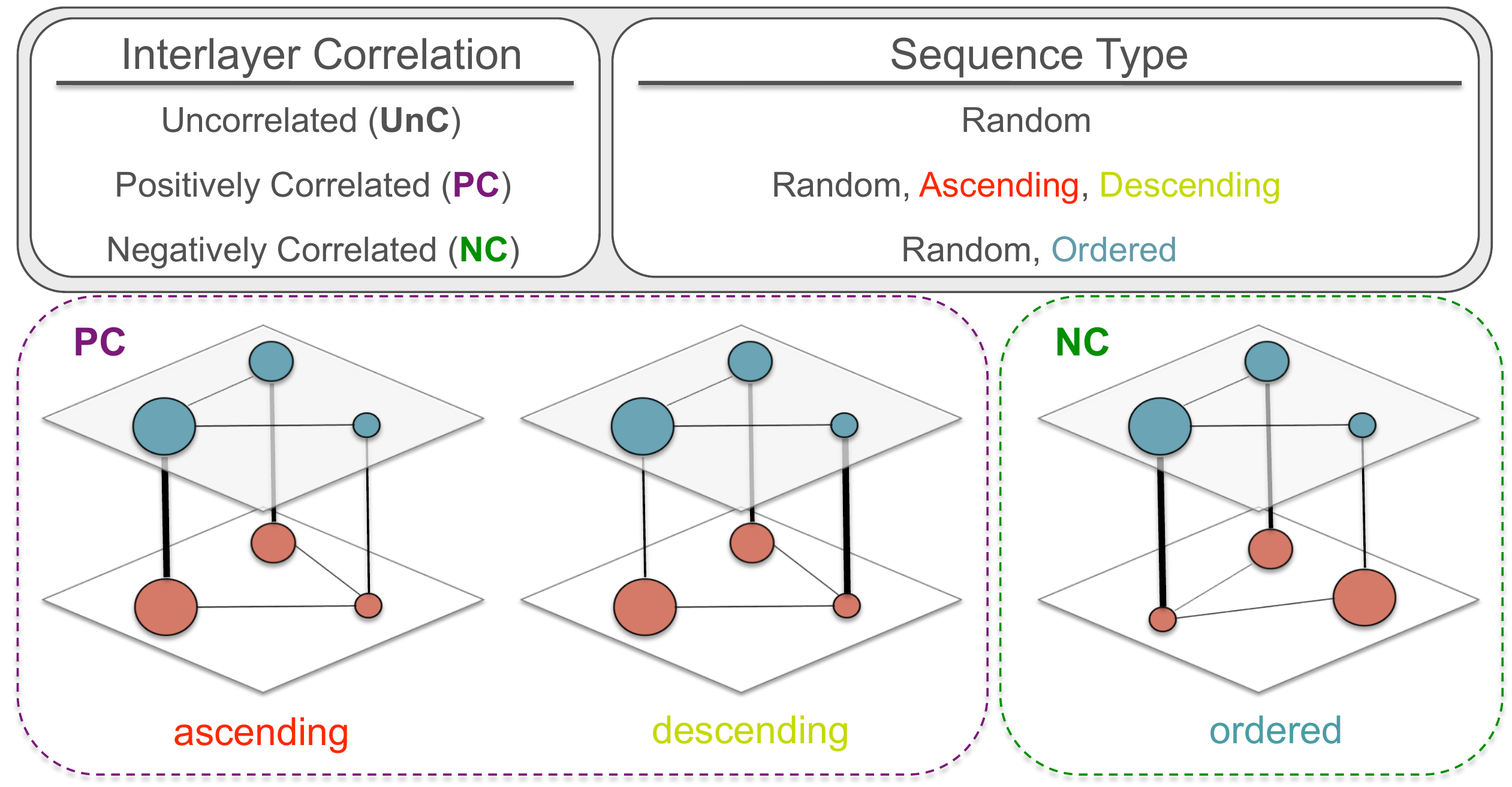}
\caption{Some connection sequence schema for a positively-correlated and negatively-correlated multiplex. \label{conschemas}}
\end{figure}

The first step is to describe how the networks used in the analysis are built. We consider multiplexes formed of statistically equivalent networks with the same number of nodes $N$ on each layer. For simplicity, the multiplexes are formed of only two layers $G_1$ and $G_2$, which along with the inter-layer connections constitute the multilayer $G$. In each of the layers, the network is built using a Molloy-Reed configurational algorithm with $\gamma = 2.5$ \cite{Molloy1995} and a weight of unity on each intra-layer link. The inter-layer edges connect a fraction $q$ of the $N$ nodes of the two layers with an intensity $p$. Three procedures to draw inter-layer connections are considered depending on the inlayer number of connection of the nodes (degree) (see Figure \ref{conschemas}). The first one is simply uncorrelated (UnC) regardless of the nodes inlayer degrees. This is the baseline scenario and has been profusely used in the literature. The other two procedures include some correlation between the degrees of the nodes connected across layers. The positively-correlated (PC) method preferentially connects high degree nodes in one layer $G_1$ with their counterparts in the other layer $G_2$. Conversely, the negatively-correlated (NC) procedure establishes links between high degree nodes in $G_1$ and low degree nodes in $G_2$ and vice versa. Within the different categories of correlations, the inter-layer links can be drawn at random or following some order based on the nodes inlayer degrees. The possible sequence types  corresponding to each layer correlation is shown in Fig.~\ref{conschemas}. For a PC multiplex, an ascending sequence connects node-pairs starting from those that have the lowest degree, the descending sequence starts from the node-pairs with the highest degree. In a NC multiplex the ordered sequence establishes connections between high-degree nodes on one layer and the low-degree nodes in another first. Thus, to create a multiplex $M(p,q)$ we must specify a) the size $N$ of each layer, b) the layer correlation strategy (UnC, PC or NC), c) the strength of inter-layer connections $p$ and the fraction of connected nodes $q$, and d) the sequence type that gives the order in which layers become interconnected.

For each realization of $M(p,q)$, we compute the algebraic connectivity $\lambda_2$ of the supralaplacian $\mathcal{L}$. The procedure for constructing the supralaplacian of a multiplex with $q = 1$ is detailed in \cite{Radicchi2013}, yielding
\begin{equation}
\mathcal{L} = \left(
\begin{array}{cc}
\mathcal{L}_1 + p \, \mathbbm{1} & - p \mathbbm{1}\\
- p \mathbbm{1} & \mathcal{L}_{2} + p \, \mathbbm{1}
\end{array} 
\right) \, ,
\label{eq:lap}
\end{equation}
where $\mathcal{L}_{k}$ stands for the Laplacian of the layer $G_{k}$, separately, and $\mathbbm{1}$ is the unit $N \times N$ matrix. In order to implement a situation in which some nodes are not connected across layers ($q \neq 1$), we assign to such nodes, e.g. node $i$ in $G_1$, a partner down in $G_2$ (arbitrarily, the node with the same index $i$), but set the corresponding inter-layer link intensity equal to zero, $p_i = 0$. Thus, a multiplex $M(p, q)$ will have a supralaplacian where the elements in the matrices $p \, \mathbbm{1}$ corresponding to positions $i$ are zero instead of $p$.

\begin{figure}
\centering
\includegraphics[width = 8.6cm]{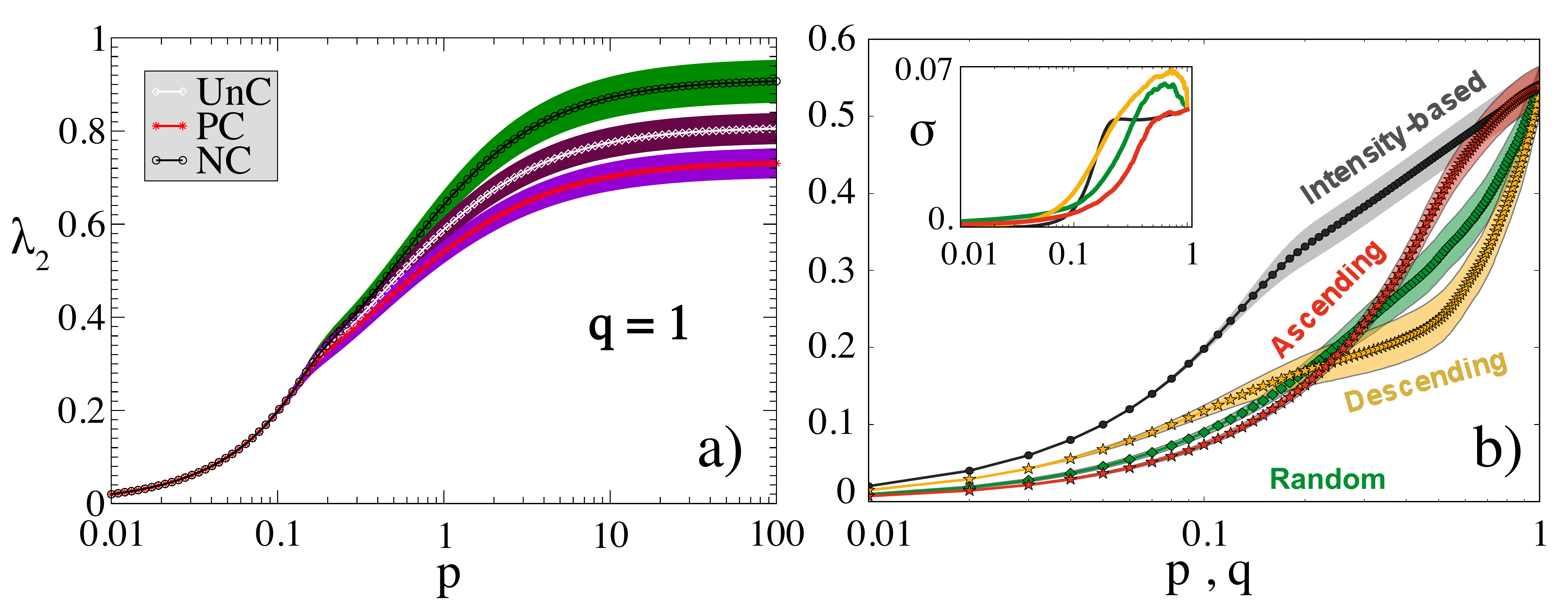}
\caption{$\lambda_2$ for a $N = 100$ scale-free multiplex. In a), mean and standard deviation over realizations (shaded region) of $\lambda_2$ of $M(p, q = 1)$ as a function of $p$ with positive (PC), negative (NC), and neutral (UnC) inter-layer correlations, for $100$ realizations. In b), average $\lambda_2$ for the PC case of a) (labelled `intensity-based') as a function of $p$, as well as the trends of $\lambda_2$ as a function of $q$ for multiplexes $M(p = 1, q)$ with PC layers and three different connection sequences, all for $10^3$ realizations. Inset: standard deviation of the respective curves across the realizations. \label{distribution}}
\end{figure}

The behavior of $\lambda_2$ for different multiplex structures $M(p,q)$ is shown in Figure \ref{distribution} as a function of the inter-layer link intensity $p$. In Figure  \ref{distribution}a, the multiplex is complete and we recover the results of \cite{Radicchi2013}. The algebraic connectivity displays an inflection point for a particular value of $p$, $p_x$, around $0.2$ regardless of the correlations between the degrees of the nodes in both layers. This point marks the beginning of a different type of dynamics, passing from a single-layer driving one for $p < p_x$ to an integrated network one for larger $p$. The correlations do play a role in the second integrated regime where they influence the relaxation time of the system (slower for PC and faster for NC with respect to the uncorrelated baseline). Increased system size $N$ only magnifies these differences. Interestingly, after the inflection point the fluctuations in $\lambda_2$ between realizations of the multiplex increase and remain constant with increasing $p$. The particular position of $p_x$ displays a dependency on the system size in the range explored here numerically, but the presence of this regime of high $\lambda_2$ variability across realizations always appears. It is worth noting as well that in contrast to ordinary phase transitions the increase in $\sigma$ is not constrained to the neighborhood of $p_x$. 

If the intensity of the inter-layer connections is kept constant, $p = 1$, and $q$ is varied instead, these curves notably change and their shape depends on the particular order implemented in the inter-layer node connection to built the multiplex (see Figure \ref{distribution}b). $\lambda_2$ is lower for a multiplex with fewer inter-layer links (with heavier weight of unity) than for a multiplex with fully interconnected layers but at small intensities. That is to some extent intuitive, as the missing inter-layer connections might have been much more necessary for the `emergent' multiplex, and  having them present albeit at a small intensity would connect the layer more strongly. In addition, the three possible sequences behave differently depending on $q$. For small $q$, until about $q < q = p_x$, they increase linearly with $q$. The descending sequence results in a consistently greater $\lambda_2$, implying a faster dynamics on the multiplex (this seems to be a generic result \cite{Aguirre2014}). This is manifestly not the case at high $q$ where the situation reverses. Besides the average $\lambda_2$, $\sigma$ displays a similar behavior as in the case of $q = 1$: beyond $p_x$ it increases and remains high afterward, although with peculiarities due to the sequence and correlations of the inter-layer connections (see inset of Figure \ref{distribution}b).

\begin{figure}
\centering
\includegraphics[width = 8.6cm]{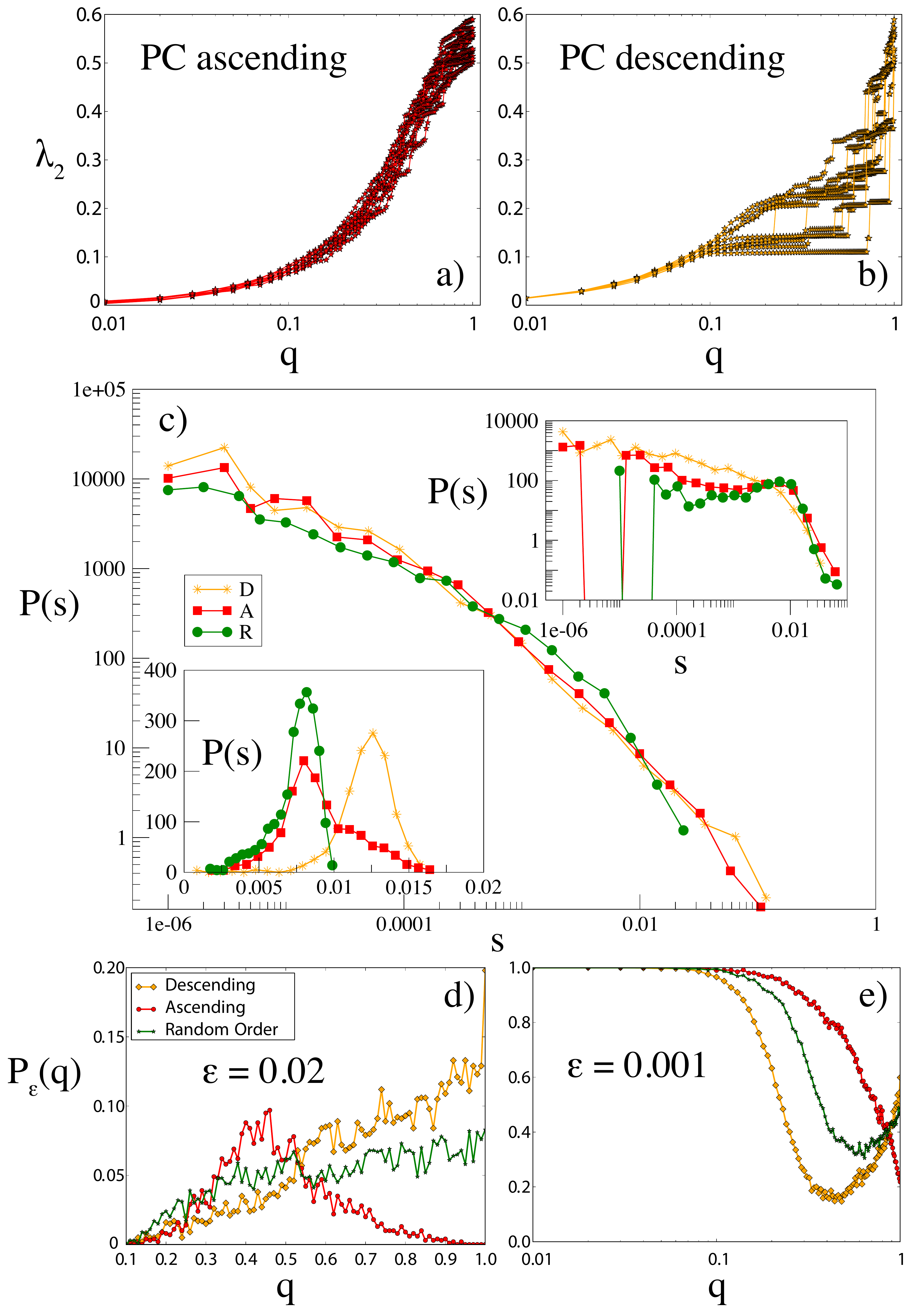}
\caption{Partial Connectivity Sequences in positively-correlated SF multiplex with $N = 100$, $p = 1$. In a) and b), $\lambda_2$ as a function of $q$ for the ascending and descending degree sequences. The plots contain ten realization in total although the curves are displayed realization by realization. In c), probability density distribution of the differences between consecutive
values of $\lambda_2$ as a new link is introduced for the three difference sequence types D (descending), A (ascending) and  R (random), computed over $10^3$ realizations at $q = 0.9$ (main plot), $q = 0.2$ (top 
right inset) and $q = 0.05$ (bottom left inset). Main plot and right inset are log-log scale, left inset is linear scale. In d) and e), probability $P_{\epsilon}$ to see jumps higher  than $\epsilon$, computed at each $q$ value over $1000$ realizations, for each of the ascending, descending and neutral sequences. \label{jumps}}
\end{figure}

On single multiplex realizations instead of on average properties, we find what is the most important result of this work. The multiplex $M(p = 1, q)$ from the previous section is constructed by adding inter-layer links of weight unit one by one until their fraction is equal to $q$. This process accumulates microscopic changes to end up in a macroscopic configuration of the multilayer networks. Naively, one could expect one of such microscopic changes to be innocuous to the macroscopic picture and, consequently, to have a very minor impact on the dynamics of any process taking place on the multiplex. However, this expectation is wrong as can be seen in Figure \ref{jumps}a and \ref{jumps}b on an example with a positively correlated multiplex. The value of $\lambda_2$ experiences significant jumps after the introduction of a single link across layers. The smooth behavior characteristic of the absence of jumps in $\lambda_2$ is only visible in a 
sparsely interconnected multiplex (lower $q$), where for some range (but 
 depending on the sequence) the increase is linear, $\lambda_2(q) \propto q$.
Although the frequency and location of $\lambda_2(q)$ discontinuities depend on the specific realization, on average they happen at some large $q$. Since $\lambda_2$ is a global feature of the multiplex, the `offending' link is intrinsically connected to
some global property of the layers, or to the sequences
themselves.  This reflects the precarious nature of adding final inter-layer
connections, implying that after some minimal number of connections has been 
set, it becomes very hard to predict the exact effect of the addition of each 
inter-layer link.

\begin{figure*}
\centering
\includegraphics[width = 17.2cm]{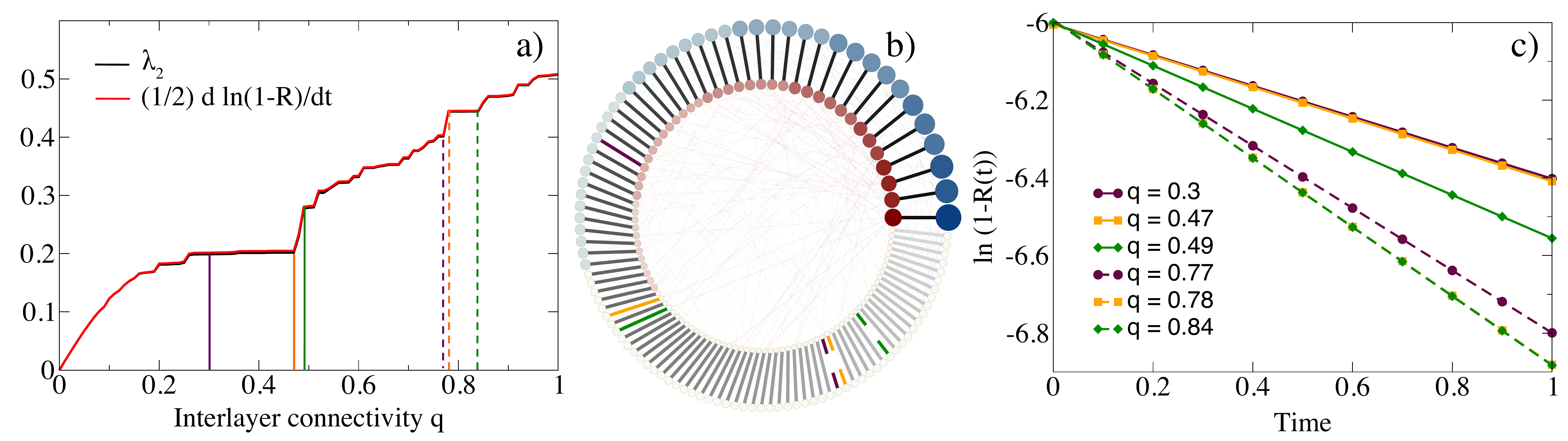}
\caption{Validation of the effects of the $\lambda_2$ jumps on the timescales of processes running on a PC multiplex $M(p = 1, q)$ with $N = 100$ generated with a sequence descending in degree. In a), $\lambda_2$ and the numeric derivative of $ln (1-R(t))$ as a function of $q$. In b), a graphical representation  of the multiplex $M(p = 1,q^{*})$ in concentric circles, with nodes arranged counterclockwise and their size and location
proportional to their degree, where $q^{*}$ is from the set of lines in panel a) with the colors corresponding. In c), the temporal relaxation of the order parameter $\ln(1 - R(t))$ for each $q^*$. \label{verification}}
\end{figure*}

The distribution of jumps is included in Figure \ref{jumps}c, where beyond the sequence the width of the distribution is controlled by the final value of $q$. Varying this parameter, one can observe long tailed distributions in the changes of $\lambda_2$ spanning several orders of magnitude as in the main plot or in the top-right inset, or more constrained jump distributions as those of the left-bottom inset for smaller values of $q$. For most of the values of $q$, the dominating effect is the non trivial long-tailed distributions. To explore this further, we introduce a cutoff $\epsilon$ and measure the probability $P_{\epsilon}(q)$ to have an increase in $\lambda_2$ by at least $\epsilon$ 
at the addition of the subsequent link. At intermediate $\epsilon$ values (Figure \ref{jumps}d), it 
identifies that the main difference between the ascending and the descending
sequence is not that the former has fewer jumps (which could have been 
expected from the lower standard deviation), but that the jumps are 
centered at some $q$, with a well-defined mean. The descending
sequence and the random one, on the other hand, show a completely different profile of a monotonically increasing jump probability. If the last node pairs to be interconnected
are those with the lowest degrees, their addition is much more 
likely to case a sudden jump in the speeds of the diffusion, than if the 
last interconnected nodes had high degree. That, however, is because in the 
latter case the multiplex would already have been relatively fast. The reason lies in another curious observation shown in Figure~\ref{jumps}e.
The probability of \emph{at least a  minute jump} is minimized at some $q < 1$ for the  descending and random sequences, and decreases monotonically for the ascending one. This lack of change is difficult to achieve if lower-degree
nodes were interconnected first. 

Calculating $\lambda_2$ implies non-linear operations and one might wonder if these results were not an artifact of the numerical methods. To verify their relevance for dynamical processes occurring on the multiplexes, we run the Kuramoto model and compare the variations in $\lambda_2$ with the synchronization timescales. Specifically, an oscillator is placed in each node $i$ of the multiplex with phase $\theta_i$. The oscillators have the same natural frequency and the evolution of their phases is described after linarization by the following equation \cite{YK1984, Arenas2008}   
\begin{equation}
\dot{\theta}_j = \sum\limits_{k \neq j} W_{jk}\left( \theta_k - \theta_j \right),
\label{Kura} 
\end{equation}
where $W_{jk}$ are the elements of the adjacency matrix with the corresponding weights for intra and inter-layer links. To ensure an easier measurement of the timescales, the initial state of the system is set at the value of the eigenvector associated with $\lambda_2$. Under these conditions, the system approaches synchronization as
\begin{equation}
1 - R(t) \propto e^{-2 \lambda_2 t},
\end{equation}
where $R$ is the phase coherence of the population, $R = |\sum_{j} e^{i \theta_j}|$, and plays the order parameter in the synchronization transition.  

In Figure \ref{verification}, we show how the changes in the spectral gap $\lambda_2$ actually translate to changes in the relaxation times. In Figure \ref{verification}a, $\lambda_2$ and the inverse relaxation times are displayed as a function of $q$ for a PC multiplex built in descending degree sequence. Both coincide as it must be if the numeric estimation of $\lambda_2$ is appropriate. Several jumps can be observed and are marked with vertical lines of different colors and textures. These changes occur after the introduction of single links as represented in Figure \ref{verification}b. The presence of these single links brings about an important variation in the macroscopic system relaxation as can be seen in Figure \ref{verification}c. The underlying exponential decays for a select range of $q$ values chosen to lie around two significant jumps. The observed straight lines imply that 
$\alpha$ is well-defined as an exponent even in the vicinity of the jumps, and hence that the 
decrease in timescales is indeed abrupt. 

Our results thus show that the jumps in the algebraic connectivity are not merely a numeric artifact, but instead correspond to the measured abrupt decreases in the synchronization timescales of the dynamical system. This ratifies that microscopic changes in the topological properties of these multilayer networks lead to effects noticed at a global scale. We also investigated the characteristics of these offending node-pairs
whose interconnection causes the jumps, but did not detect any special local network feature. The lack of any
correspondence between the degree, and the degree of the neighbors, of the select nodes points to the 
nontrivial nature of the connection between the network properties of the multiplex nodes and their role as
crucial mediators in enhancing the synchronization time of the multiplex, and calls for the use of global spectral methods to determine their location and quantify the possible effects of their introduction or deletion.

Partial financial support has been received from the Spanish Ministry
of Economy (MINECO) and FEDER (EU) under the project ESOTECOS
(FIS2015-63628-C2-2-R), from the EPSRC project GALE, EP/K020633/1 and from the EU Commission through project LASAGNE (318132). JJR acknowledges funding from the Ram\'on y Cajal program of MINECO.


\begin{thebibliography}{45}

\bibitem{Boguna2003}
M. Bogu\~n\'a, R. Pastor-Satorras,  and A. Vespignani, in  {\emph Statistical Mechanics of Complex Networks}, Lecture Notes in Physics, Vol. 625, edited by R. Pastor-Satorras, M. Rubi, and A. D\'{\i}az-Guilera (Springer, Heidelberg, 2003) pp. 127--147.

\bibitem{Jamakovic2006}
A. Jamakovic, R. Kooij, P. van Mieghem, and E. van Dam, Proceedings of the 13th Annual Symposium on Communications and Vehicular Technology in the Benelux , 35 (2006).

\bibitem{Dorogovtsev2008}
S.~N. Dorogovtsev, A.~V. Goltsev, and J.~F.~F. Mendes, Rev. Mod. Phys. {\bf 80}, 1275 (2008).

\bibitem{Atay2006}
F.~M. Atay, T. Bitikoglu, and J. Jost, Physica D: Nonlinear Phenomena {\bf 224}, 35 (2006).

\bibitem{Arenas2006}
A. Arenas, A. D\'iaz-Guilera, and C.~J. P\'erez-Vicente, Physica D: Nonlinear Phenomena {\bf 224}, 27 (2006).

\bibitem{GG2007}
J. G\'omez-Garde\~nes, Y. Moreno, and A. Arenas, Phys. Rev. Lett. {\bf 98}, 034101 (2007).

\bibitem{Chen2012}
J. Chen, J.-A. Lu, C. Zhan, and G. Chen, in {\it Handbook of Optimization in Complex Networks}, Springer Optimization and Its Applications, Vol. 57, edited by M.~T. Thai and P.~M. Pardalos (Springer US, 2012) pp. 81--113.

\bibitem{Almendral2007}
J.~A. Almendral and A. D\'{i}az-Guilera, New Journal of Physics {\bf 9}, 187 (2007).

\bibitem{Estrada2015}
E. Estrada, E. Vargas-Estrada, and H. Ando, Phys. Rev. E {\bf 92}, 052809 (2015).
  
\bibitem{Hernandez2014}
J. Mart\'in-Hern\'andez, H. Wang, P. Van~Mieghem, and G. D'Agostino, Physica A: Statistical Mechanics and its Applications {\bf 404}, 92 (2014).
  
\bibitem{Lovasz1994}
L. Lov\'asz, Tech. Report YALEU/DCS/TR-1029, Department of Computer Science, Yale University (1994).

\bibitem{Boccaletti14}
S. Boccaletti, G. Bianconi, R. Criado, C. del Genio, J. G\'omez-Garde{\~n}es,  I. Sendi{\~n}a-Nadal, Z. Wang, and M. Zanin, Physics Reports {\bf 544}, 1 (2014). 

\bibitem{Kivela14}
M. Kivel\"a, A. Arenas, M. Barthelemy, J.~P. Gleeson, Y. Moreno, and M.~A. Porter, Journal of Complex Networks {\bf 3}, 203 (2014).

\bibitem{Goh2015}
K.-M. Lee, B. Min, and K.-I. Goh, The European Physical Journal B {\bf 88}, 1 (2015).

\bibitem{Iasio2013}
G. di~Iasio, S. Battiston, L. Infante, and F. Pierobon, Munich Personal RePEc Archive, paper 52141, available at https://mpra.ub.uni-muenchen.de/52141 (2013).

 
\bibitem{Thurner2015}
S. Poledna, J.~L. Molina-Borboa, S. Mart\'{\i}nez-Jaramillo, M. van der Leij, and S. Thurner, Journal of Financial Stability {\bf 20}, 70 (2015).

\bibitem{Burkholz2015}
R. Burkholz, M.~V. Leduc, A. Garas, and F. Schweitzer, available at  http://arxiv.org/abs/arXiv:1506.06664 (2015).

\bibitem{Domenico2014}
M. De~Domenico, A. Sol\'e-Ribalta, S. G\'omez, and A. Arenas, Procs.
  Natl. Acad. Sci. U.S.A. {\bf 111}, 8351 (2014).
  
\bibitem{Strano2015}
E. Strano, S. Shai, S. Dobson, and M. Barthelemy, Journal of The Royal Society Interface {\bf 12}, 20150651 (2015).
  
\bibitem{Granell2013}
C. Granell, S. G\'omez, and A. Arenas, Phys. Rev. Lett. {\bf 111}, 128701 (2013).

\bibitem{Vazquez2015}
 F. Vazquez, M.~A. Serrano, and M. San~Miguel, available at
http://arxiv.org/abs/arXiv:1511.05606 (2015).

\bibitem{Mucha10}
P.~J. Mucha, T. Richardson, K. Macon, M.~A. Porter, and J.-P. Onnela, Science {\bf 328} (2010).

\bibitem{Buldyrev2010}
S.~V. Buldyrev, R. Parshani, G. Paul, H.~E. Stanley, and S. Havlin, Nature {\bf 464}, 1025 (2010). 

\bibitem{Klimek2013}
P. Klimek and S. Thurner, New Journal of Physics {\bf15}, 063008 (2013).

\bibitem{Csermely2013}
P. Csermely, Talent Dev. and Excellence {\bf 5}, 115 (2013).

\bibitem{Wang2013}
Z. Wang and A.~S.~M. Perc, Sci. Rep. {\bf 3}, 2470 (2013).

\bibitem{Diakonova2014}
M. Diakonova, M. San~Miguel, and V.~M. Egu\'iluz, Phys. Rev. E {\bf 89}, 062818 (2014).

\bibitem{Klimek2015}
P. Klimek, M. Diakonova, V.~M. Egu\'iluz, M. San Miguel, and S. Thurner, available at http://arxiv.org/abs/arXiv:1601.01576 (2016).

\bibitem{Diakonova2015}
M. Diakonova, V. Nicosia, V. Latora, and M. San Miguel, New J. Phys. {\bf 18}, 023010 (2016).
  
\bibitem{Gomez2013}
S. G\'omez, A. D\'{i}az-Guilera, J. G\'omez-Garde\~nes, C.~J. P\'erez-Vicente, Y. Moreno, and A. Arenas, Phys. Rev. Lett. {\bf 110}, 028701 (2013).

\bibitem{SoleRibalta2013}
A. Sol\'e-Ribalta, M. De~Domenico, N.~E. Kouvaris,  A. D\'{\i}az-Guilera, S. G\'omez, and A. Arenas, Phys. Rev. E {\bf 88}, 032807 (2013).

\bibitem{Sahneh2015}
F. Darabi~Sahneh, C. Scoglio, and P. Van Mieghem, Phys. Rev. E {\bf 92}, 040801 (2015). 

\bibitem{Radicchi2013}
F. Radicchi and A. Arenas, Nature Physics {\bf 9}, 717 (2013).

\bibitem{Garrahan2014}
I.~L. Juan P.~Garrahan, available at http://arxiv.org/abs/arXiv:1406.4706 (2014). 

\bibitem{Bastas2015}
N. Bastas, F. Lazaridis, P. Argyrakis, and M. Maragakis, EPL (Europhysics Letters) {\bf 109}, 38006 (2015).

\bibitem{Shakeri2015}
H. Shakeri, N. Albin, F. Darabi~Sahneh, P. Poggi-Corradini, and C. Scoglio, Phys. Rev. E {\bf 93}, 030301 (2016).

\bibitem{Gambuzza2015}
L.~V. Gambuzza, M. Frasca, and J. G\'omez-Garde\~nes, EPL (Europhysics Letters) {\bf 110}, 20010 (2015).

\bibitem{Radicchi2014}
F. Radicchi, Phys. Rev. X {\bf 4}, 021014 (2014).  

\bibitem{Enzo2015}
V. Nicosia and V. Latora, Phys. Rev. E {\bf 92}, 032805 (2015).

\bibitem{Molloy1995}
M. Molloy and B. Reed, Random Structures \& Algorithms {\bf 6}, 161 (1995).

\bibitem{Aguirre2014}
J. Aguirre, R. Sevilla-Escoboza, R. Guti\'errez, D. Papo, and J.~M. Buld\'u, Phys. Rev. Lett. {\bf 112}, 248701 (2014).

\bibitem{YK1984}
Y. Kuramoto, {\emph Chemical Oscillations, Waves, and Turbulence} (Springer--Verlag, New York, 1984).

\bibitem{Arenas2008}
A. Arenas, A. D\'{\i}az-Guilera, J. Kurths, Y. Moreno, and C. Zhou, Physics Reports {\bf 469}, 93 (2008).

\end{thebibliography}

\end{document}